\documentclass[10pt, conference, letterpaper]{IEEEtran}
\IEEEoverridecommandlockouts
\usepackage{graphicx}
\usepackage{balance}
\usepackage{comment}

\usepackage{booktabs}
\newcommand{\ra}[1]{\renewcommand{\arraystretch}{#1}}

\usepackage[ruled]{algorithm2e}
\usepackage{verbatim} 
\usepackage{graphics} 
\usepackage{epsfig} 
\usepackage{epstopdf}
\usepackage{subfigure}
\usepackage{balance}
\usepackage{comment}
\usepackage{booktabs}
\usepackage{amsmath}
\usepackage{amssymb}
\usepackage{amsfonts}
\usepackage{bm}
\usepackage{colortbl}
\usepackage{tabularx}
\usepackage{color}
\usepackage{xspace}
\usepackage{graphicx}    
\usepackage{url}    
\usepackage{multirow}

\usepackage{etoolbox}
\makeatletter
\patchcmd{\@makecaption}
  {\scshape}
  {}
  {}
  {}
\makeatother

\usepackage{cite}
\usepackage{amsmath,amssymb,amsfonts}
\usepackage{graphicx}
\usepackage{textcomp}
\usepackage{xcolor}
\def\BibTeX{{\rm B\kern-.05em{\sc i\kern-.025em b}\kern-.08em
    T\kern-.1667em\lower.7ex\hbox{E}\kern-.125emX}}
\begin{document}

\title{Skin-MIMO: Vibration-based MIMO Communication over Human Skin}
\author{Dong Ma, \textit{Student Member,~IEEE,} Yuezhong Wu, \textit{Student Member,~IEEE,} Ming Ding, \textit{Senior Member,~IEEE,} \\ Mahbub Hassan, \textit{Senior Member,~IEEE,} and Wen Hu, \textit{Senior Member,~IEEE,}
\thanks{Dong Ma, Yuezhong Wu, Mahbub Hassan, and Wen Hu are with the School of Computer Science and Engineering, University of New South Wales (UNSW), Sydney, Australia (E-mail: dong.ma1@unsw.edu.au, yuezhong.wu@student.unsw.edu.au, mahbub.hassan@unsw.edu.au, wen.hu@ unsw.edu.au).}
\thanks{Ming Ding is with Data61, CSIRO, Eveleigh, Australia (E-mail: ming.ding@data61.csiro.au). }

}


\maketitle

\begin{abstract}
We explore the feasibility of Multiple-Input-Multiple-Output (MIMO) communication through vibrations over human skin. Using off-the-shelf motors and piezo transducers as vibration transmitters and receivers, respectively, we build a 2x2 MIMO testbed to collect and analyze vibration signals from real subjects. Our analysis reveals that there exist multiple independent vibration channels between a pair of transmitter and receiver, confirming the feasibility of MIMO. Unfortunately, the slow ramping of mechanical motors and rapidly changing skin channels make it impractical for conventional channel sounding based channel state information (CSI) acquisition, which is critical for achieving MIMO capacity gains. To solve this problem, we propose Skin-MIMO, a deep learning based CSI acquisition technique to accurately predict CSI entirely based on inertial sensor (accelerometer and gyroscope) measurements at the transmitter, thus obviating the need for channel sounding. Based on experimental vibration data, we show that Skin-MIMO can improve MIMO capacity by a factor of 2.3 compared to Single-Input-Single-Output (SISO) or open-loop MIMO, which do not have access to CSI. A surprising finding is that gyroscope, which measures the angular velocity, is found to be superior in predicting skin vibrations than accelerometer, which measures linear acceleration and used widely in previous research for vibration communications over solid objects. 
\end{abstract}


\section{Introduction}

We are at the cusp of a revolution in wearable computing. 
In the near future, 
people are expected to wear many types of sensors and devices on their bodies for better health, comfort, entertainment, and convenience. 
A recent survey~\cite{seneviratne2017survey} has revealed that the market is already beaming with hundreds of different types of smart wearable products including smartwatches, wrist bands, smart glasses, smart headsets, smart jewellery, electronic garments, skin patches, and so on. 
When multiple devices are worn on the same body, 
they are expected to form a new type of communication network, 
called Body Area Network (BAN)~\cite{movassaghi2014wireless}, 
to further increase the capability and utility of wearable computing.     

While current standards consider radio frequency as the communication medium for BAN, 
\textit{skin vibration} remains an untapped and underexplored alternative for realizing communications between two wearable devices attached at different parts of the body. 
For BAN, 
enabling a vibration communication channel promises several complementary advantages compared to radio-based communications as it can reduce the risk of eavesdropping for sensitive body-related data, avoid severe radio interference in crowded places, 
and improve usability in radio-hazard environments such as airborne vehicles. 
As many consumer wearable products already include motors for haptic purposes~\cite{nadi,haptic,thar2017haptictoolkit}, transmitting vibration signals over the skin can be readily achieved. 
Similarly, 
vibration sensors such as accelerometers are also included in many wearable devices, 
which can be used to detect and monitor vibration signals~\cite{roy2015ripple,hwang2012privacy,yonezawa2015vinteraction,kim2015vibration}. 

With more vibration sensors, 
such as piezoelectric transducers and contact microphones, 
becoming widely available, 
vibration-based communication is attracting increasing attention from research community. 
By attaching a motor to the wrist and an accelerometer to the elbow in varying distance, 
Zhang et al.~\cite{zhang2017bioacoustics} analyzed frequency response and path loss of vibration signals and confirmed the feasibility of using skin vibration as a practical transmission medium for BAN. 
Roy et al.~\cite{roy2016ripple} designed and implemented an orthogonal frequency-division multiplexing (OFDM) protocol for vibration communications and reported 7.4 kbit/s data rate for a motor-fitted smart ring transmitting information through the finger to a touched object equipped with a microphone acting as the receiver. 

\begin{figure}[t]
	\centering
	\includegraphics[scale=0.6]{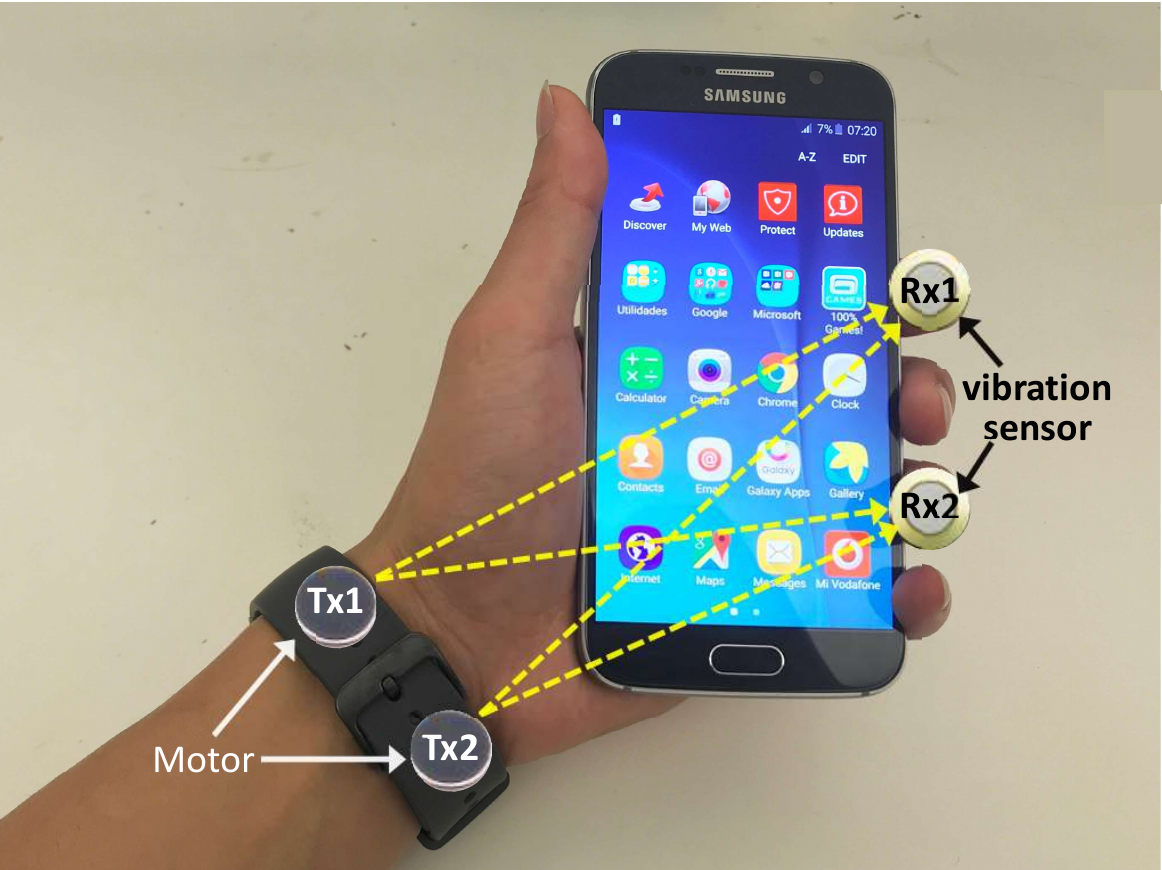}
	\caption{An illustration of body-mediated vibration MIMO communication between a smartwatch and a smartphone held in hand. }
	\label{fig:mimo_phone}
\end{figure}

While existing works have clearly demonstrated the viability of skin vibration as a practical communication medium for wearable devices, 
they studied only a single transmitter sending data to a single receiver, 
which refers to Single-Input-Single-Output (SISO) communication system. 
Whether it is possible to achieve Multiple-Input-Multiple-Output (MIMO) communication over skin vibrations remains unclear and unexplored. In this paper, 
we are motivated to explore MIMO because it has the potential to further improve the reliability and capacity of vibration communications in BAN by taking advantage of multiple motors available in emerging wearable devices\footnote{For example, some new generation smartwatches \cite{haptic} include 4 motors, one in each corner of the device to help navigate the wearer via advanced haptics.}. 
An example of vibration MIMO communication between a smartwatch and a smartphone is illustrated in Figure~\ref{fig:mimo_phone}, 
which assumes that both the smartwatch and the edge of the smartphone are equipped with multiple motors and vibration sensors.

In our pursuit of MIMO over skin vibrations, 
we collect vibration signal data from the hands of real subjects using a 2-transmitter-2-receiver prototype that we built using consumer motors and piezo transducers. 
Analysis of the vibration signal data reveals two important results not previously reported in the literature. 
First, the analysis of the 2x2 vibration channel matrix shows that the spatial channels are clearly independent and uncorrelated, 
which confirms MIMO is achievable over skin vibrations. 
The potential for vibration MIMO is created by the complex structure of bones and muscles in human bodies. 
Unfortunately, 
the second finding suggests that the ramping and ringing effect of the vibration motor, together with the short channel coherence time of human body channel, make conventional CSI acquisition through channel sounding impractical. 
Note that issue of short channel coherence time also arises in existing wireless communications when the mobile user moves so fast that the channel changes before the channel sounding can be completed. 
MIMO for such cases is then obtained without CSI acquisition, 
often referred to as open-loop MIMO, 
albeit at the expense of reduced MIMO gain. 


A fundamental contribution of the proposed Skin-MIMO protocol is to obtain CSI for vibration channels without transmitting channel sounding packets, 
i.e., harvest the gain of CSI-based closed-loop MIMO while physically operating as an open-loop one. 
We achieve this by employing deep learning to learn vibration propagation from one skin location to another. 
More specifically, 
we use the inertial sensor measurements observed at the transmitters to learn and classify the quantized CSI values observed at the receivers. 
Such CSI learning is possible because the factors, such as muscle type, bone structure, blood flow, heartbeat pattern, etc. that affect the skin vibration channel is contained within the same human body. 
Note that such learning generally would not work for conventional wireless communications as the environmental factors that affect the channel are not contained within a closed system. 

\begin{figure}[t]
	\centering
	\subfigure[]{
	\includegraphics[scale=0.5]{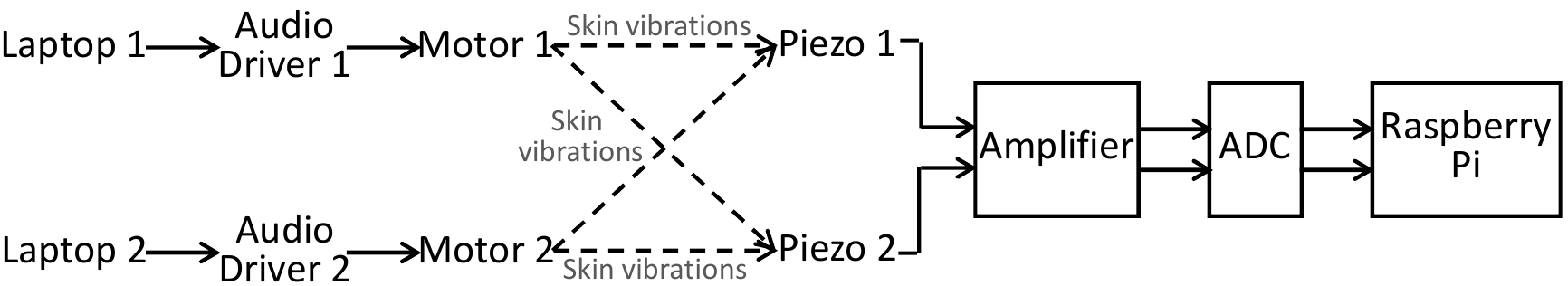}
	\label{fig:workflow}}
	\subfigure[]{
	\includegraphics[scale=0.5]{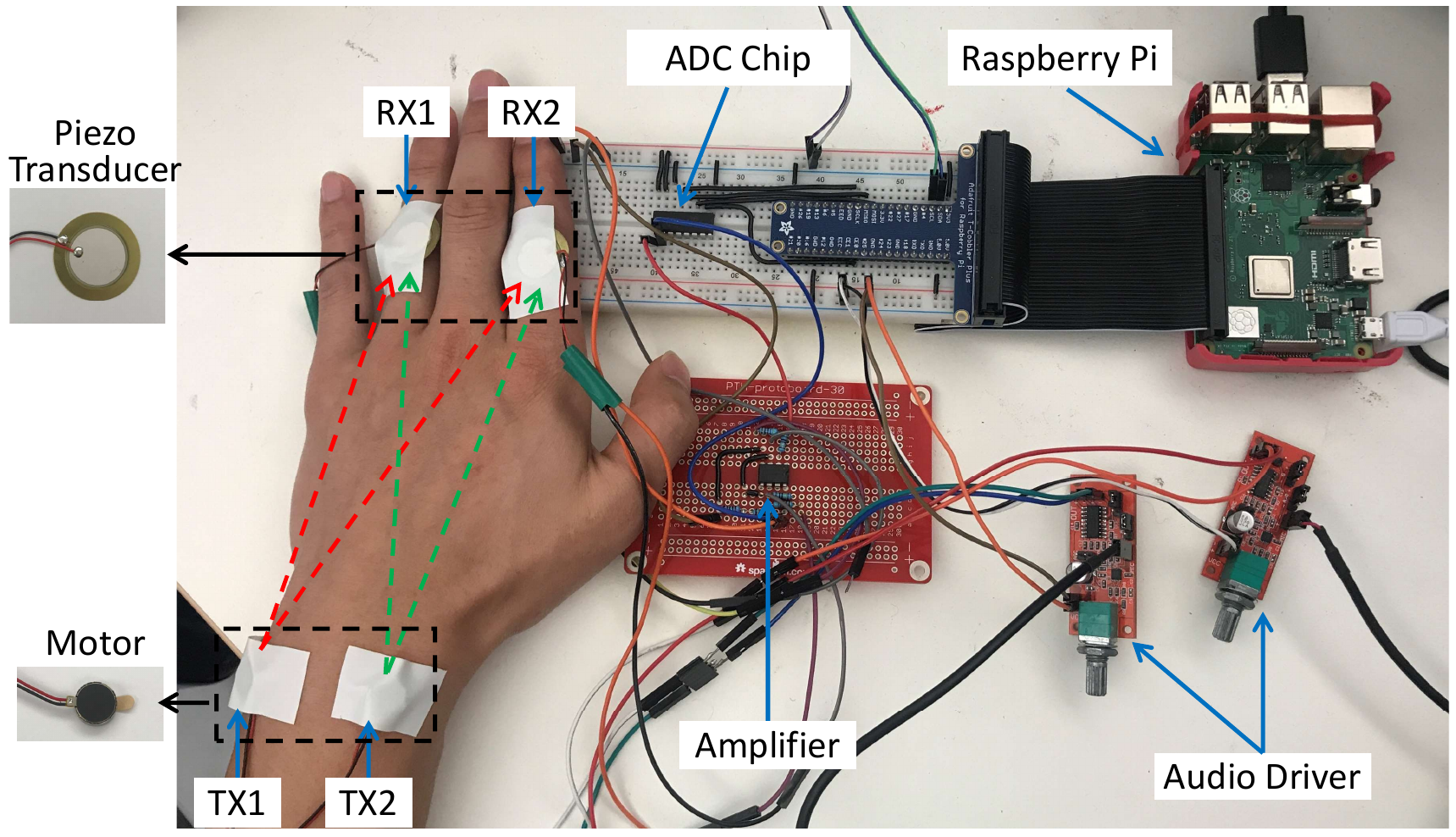}
	}
	\caption{(a) The workflow diagram of the proposed $2 \times 2$ vibration MIMO system, (b) the designed vibration MIMO testbed.}
	\label{fig:testbed}
\end{figure}

Contributions of this paper can be summarized as follows:
\begin{itemize}
    \item Using experiments with real subjects, we demonstrate that the complex bone and muscle structure of the human body can create uncorrelated vibration channels enabling vibration-based MIMO communication over human skin. (Section \ref{s:feasible}).
    
    \item We reveal that, due to long start-stop lead time of mechanical vibration devices combined with short channel coherence time of the skin, CSI acquisition using conventional channel sounding becomes impractical for vibration MIMO (Section \ref{s:coherence}).
    
    \item We propose a novel and practical deep learning-based CSI acquisition framework to learn and classify the quantized CSI values observed at the receivers simply by observing the inertial sensor (accelerometer and/or gyroscope) measurements recorded at the transmitters (Section \ref{s:deep}).

    \item We evaluate Skin-MIMO with vibration data collected from real subjects and show that use of predicted CSI increases MIMO capacity by a factor 2.3 compared to SISO or open-loop MIMO that does not have access to CSI (Section \ref{s:evaluation}).
    
    \item Finally, we discover that although accelerometer has been used widely in previous research for vibration communications over solid objects, gyroscope is actually a superior predictor of skin vibrations (Section \ref{s:evaluation}). 
\end{itemize}


\begin{figure*}[t]
	\centering
	\subfigure[subject 1]{
	    \includegraphics[scale=0.5]{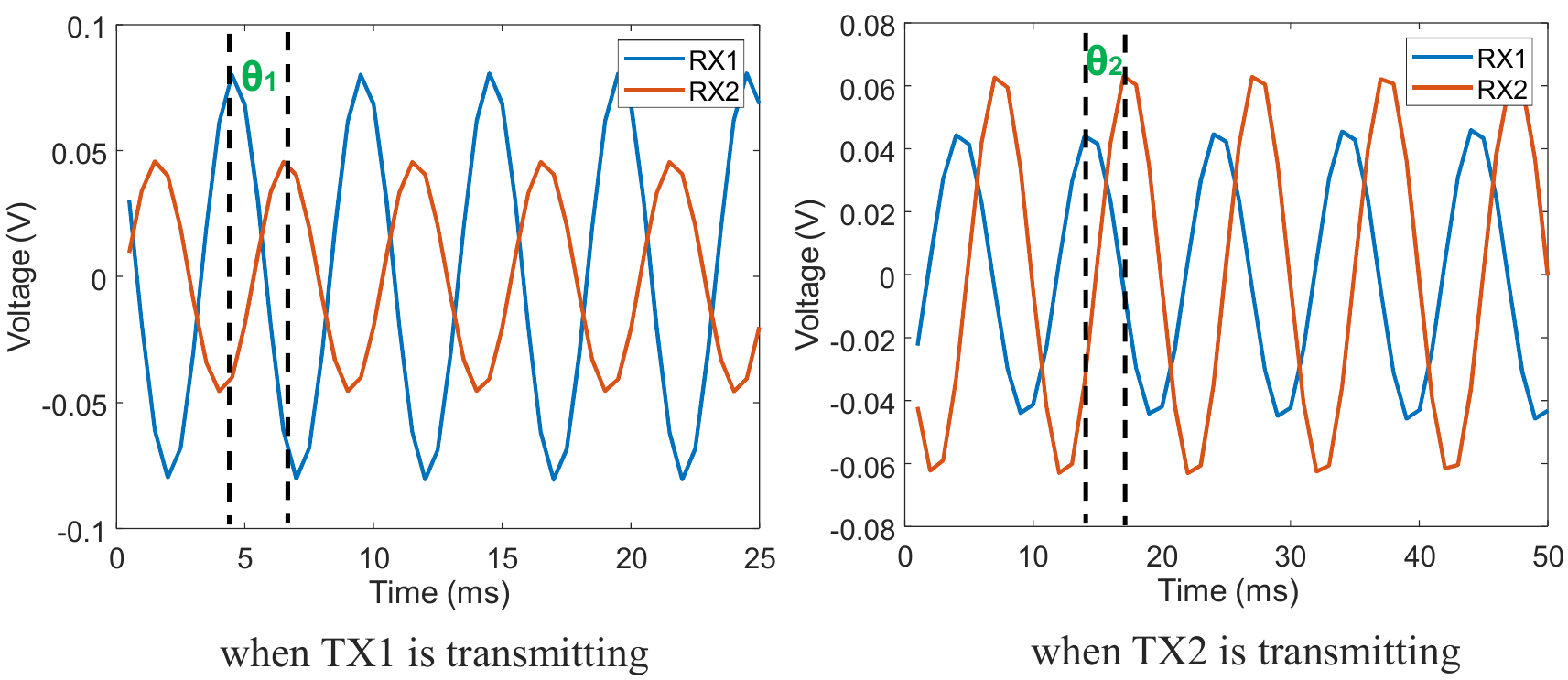}}
	\subfigure[subject 2]{
	\includegraphics[scale=0.5]{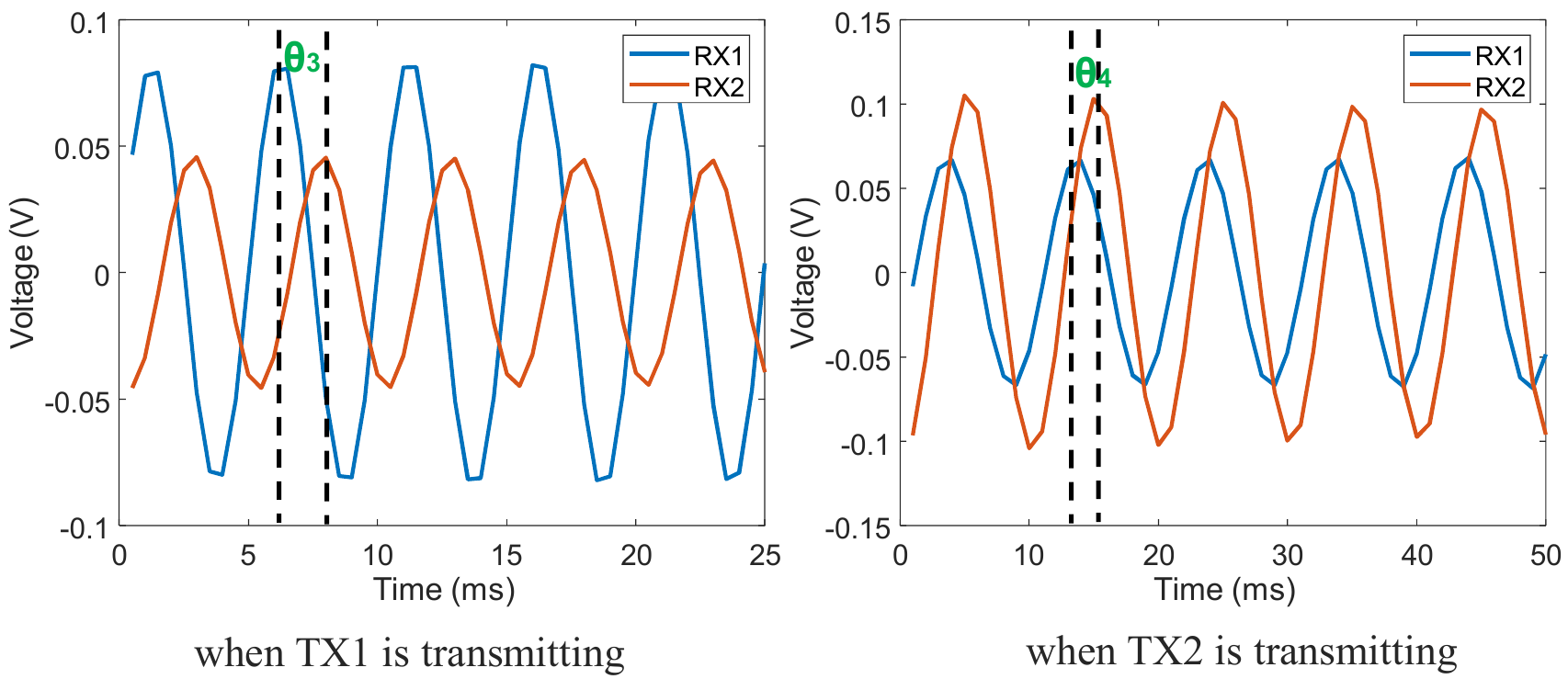}}
	\caption{Bandpass filtered signals at the RX from two subjects when activating TX1 (left) and TX2 (right), respectively, during which the other motor is muted. The exhibited amplitude and phase difference demonstrates the feasibility of vibration MIMO over human skin.}
	\label{fig:mimo_feasi}
\end{figure*}

\section{MIMO Feasibility for Skin Vibration}
\label{s:feasible}

We investigate the feasibility of vibration MIMO by building a $2\times2$ MIMO testbed as shown in Figure~\ref{fig:testbed}. 
Two AC motors\footnote{The motors are ELV1030AL from NFP Motor~\cite{motor}.} attached at the wrist with a medical tape act as the transmitters, 
while two piezo transducers\footnote{Each piezo is a metal disk with a diameter of 20mm and thickness of 0.4mm~\cite{piezo}.} attached at the index and ring fingers act as the receivers. 
The frequency and strength of the generated motor vibrations on the skin are controlled by adjusting the frequency and amplitude of its input AC voltage. 
Sine waves with different frequencies are generated in MATLAB, 
which are then fed to the motors using audio drivers~\cite{audiodriver} through the audio jacks of laptops. 
Two laptops and two audio drivers are used to achieve independent vibration control of the individual motors. 
Piezo disks generate voltage when subjected to mechanical vibrations. 
An amplifier is used to amplify the weak output from piezo transducers, 
which are then picked up and sampled by an MCP3008~\cite{mcp3008} analog-to-digital (ADC) chip before being stored in the on-device memory of a Raspberry Pi.

A key requirement to achieve MIMO is that there should exist multiple independent/uncorrelated channels between the TX and RX. 
Otherwise, 
the MIMO channel will collapse to a SISO one. 
More specifically, 
given a transmitted signal, 
the \textit{amplitude} and \textit{phase} of the received signals at different reception antennas should be distinct. 
Therefore, 
we perform the following experiments with two subjects: 
(1) activate TX1 to transmit a 200Hz sine wave and measure the output at RX1 and RX2; 
(2) activate TX2 to transmit a 200Hz sine wave and measure the output at RX1 and RX2. 
During the transmission of one motor, 
the other motor is muted.

Due to the minor movement of subject's hand during data collection as well as hardware imperfection,
there exists noise in the measured signal. 
Thus, 
we apply a bandpass filter with a passband of 190-210Hz to the raw signal to extract the useful signal being transmitted. 
Figure~\ref{fig:mimo_feasi} illustrates the filtered signals for the above two experiments. 
Visually, 
we can see that the amplitudes of the two received signals are different for both experiments and both subjects. 
This is because the vibration may experience different path loss and path length when propagating to RX1 and RX2. 
In addition, there exists a phase shift of $\theta_1$ (or $\theta_3$) and $\theta_2$ (or $\theta_4$) between the received signals at RX1 and RX2 when activating TX1 and TX2, respectively, which also suggests the distinct channel characteristics between TX and RX.

Encouraged by the above observations, 
we design our testbed, 
which consists of two transmission antennas and two reception antennas, 
thereby creating four subchannels, 
i.e., \textit{TX1-RX1, TX1-RX2, TX2-RX1, TX2-RX2}. After obtaining the amplitude $A$ and phase $\theta$ of the received signal on each subchannel, 
we can calculate the channel response $h_{mn}$ between TX antenna $n$ to RX antenna $m$ through 
\begin{equation}
h_{mn} = \frac{A_{mn}}{A_0} e^{j(\theta_{mn}-\theta_0)},
\label{eq:csi}
\end{equation}
where $A_{mn}$ and $\theta_{mn}$ are the amplitude and phase of received signal from TX antenna $n$ to RX antenna $m$, respectively. 
And $A_{0}$ and $\theta_{0}$ are the amplitude and phase of the signal transmitted from TX antenna $n$, respectively. 
After computing the channel response for each subchannel, 
a $2\times2$ channel matrix can be derived for the designed MIMO system. 
As an example, 
we plot the signal from subject 1 in Figure~\ref{fig:mimo_feasi} , 
and obtain the following channel matrix:
\[
\mathbf{H}=
\begin{bmatrix} 
h_{11} & h_{12} \\
h_{21} & h_{22} 
\end{bmatrix}
=
\begin{bmatrix} 
0.0541 + 0.0712i & -0.0151-0.0369i \\
0.0473 - 0.0153i  & 0.0253+0.0529i
\end{bmatrix}   
\]

To mathematically demonstrate vibration MIMO feasibility, 
one can calculate the rank, eigenvalues, and condition number of the channel matrix. 
The rank of the channel matrix indicates the number of data streams that can be spatially multiplexed in the MIMO channel, 
i.e., at most 2 for a $2\times 2$ MIMO system. 
Eigenvalues are two positive values for a $2\times 2$ MIMO and reflect the gain of the MIMO channels. Condition number is the ratio of the two eigenvalues (the larger one divided by the smaller one) in dB. 
A well-conditioned channel matrix should have almost equivalent eigenvalues and low condition number typically around 10dB~\cite{performance2009condition}.
We calculate the three metrics for the above channel matrix as 2, (0.0032, 0.0123), 11.67$dB$, respectively, which also confirms that vibration MIMO is achievable over human hand.

\section{Impracticability of the Conventional Channel-Sounding Based CSI Acquisition}
\label{s:coherence}

In MIMO communication, 
knowledge of channel state information (CSI), 
i.e., channel matrix, is critical~\cite{shen2016high,ding2015nonorthogonal}. 
To fully exploit the spatial multiplexing gain of MIMO, 
the transmitter needs to optimize its transmission scheme, 
such as precoding matrix design and antenna power allocation, 
based on current CSI. 
For a time-varying channel, 
CSI is valid only for a certain period of time, 
refers to as channel coherence time. 
A new CSI should be acquired after the current channel coherence time passes, 
otherwise the MIMO capacity will suffer from a significant drop due to the mismatch between the transmission scheme and the new MIMO channel. 
Therefore, 
we first measure the channel coherence time for vibration over human skin.

\begin{figure}[t]
	\centering
	\includegraphics[scale=0.52]{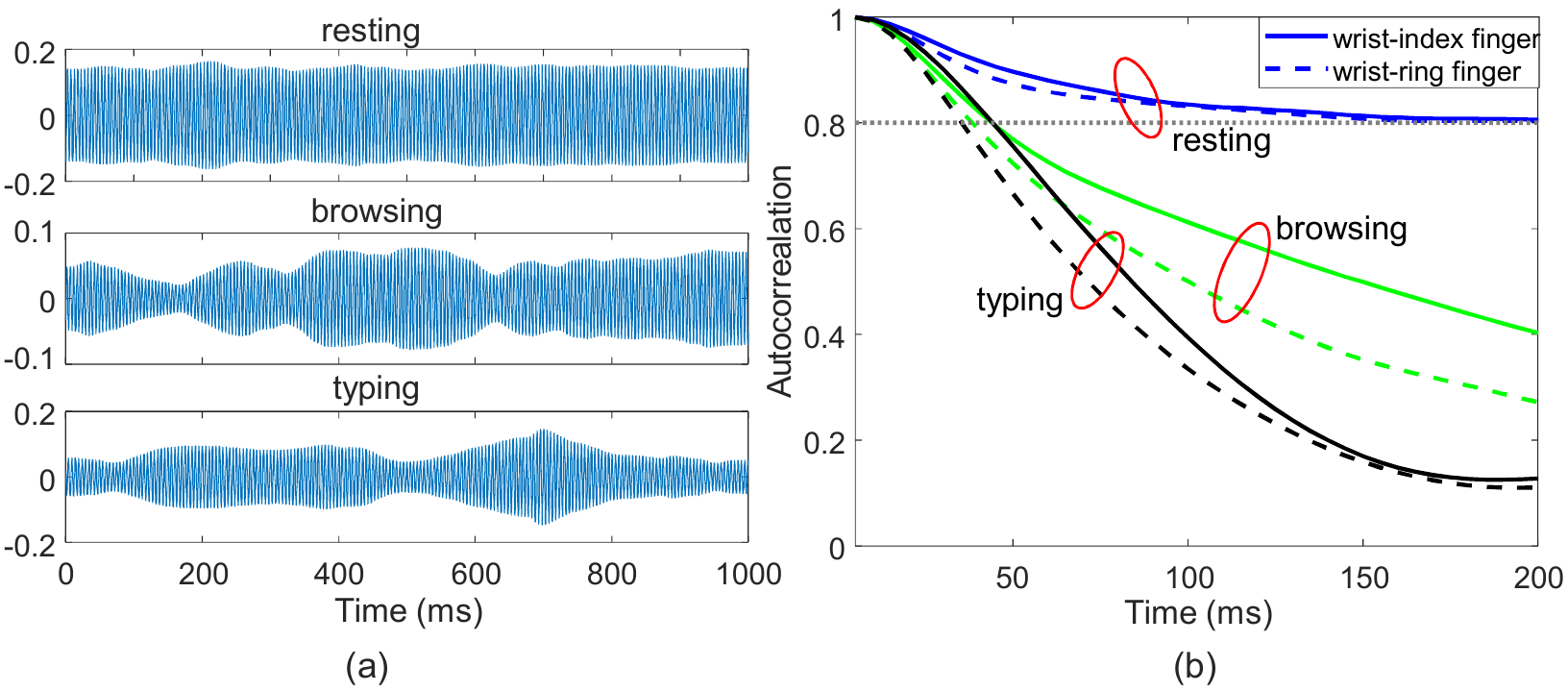}
	\caption{(a) The bandpass filtered signal at index finger and (b) its autocorrelation when transmitting signal from the wrist (TX2 only) for different activities. With correlation threshold of 0.8, the channel coherence times for resting, browsing and typing are found to be around 150ms, 44ms, and 40ms, respectively.}
	\label{fig:coherance_time}
	\vspace{-0.2in}
\end{figure}

Specifically, 
we activate TX2 (wrist) to transmit a 200Hz sine wave and measure the received signal at RX1 (ring finger) and RX2 (index finger). 
During the transmission, 
the subject holds a smartphone and performs three activities, 
i.e., resting, 
one-hand browsing, 
and one-hand typing on the screen. 
Each activity is measured for a duration of 60s.
Figure~\ref{fig:coherance_time} (a) plots a segment of the bandpass filtered signal for the three activities at index finger, 
where we can observe the amplitudes of the sine waves fluctuate over time and more intense activities (e.g., browsing and typing) incur stronger fluctuations.
Then, 
we extract the envelop of the filtered signal and calculate its time auto-correlation coefficients as illustrated in Figure~\ref{fig:coherance_time} (b). 
If we define the channel coherence time as the period when the correlation coefficient is above 0.8~\cite{paulraj2003introduction}, 
the measured channel coherence time for resting, browsing and typing is around 150ms, 44ms, and 40ms, respectively, 
which indicates that more intense activity results in shorter channel coherence time~\cite{smith2009temporal}, 
i.e., faster channel variation.

To acquire CSI for MIMO, 
conventionally, the transmitter will send a sounding packet to probe the channel. 
The receiver then calculates the CSI based on the received sounding signal and feedback the CSI to the transmitter. 
On one hand, 
each sounding packet should not be too short, which usually consists of multiple sine cycles (e.g., more than 10,000 in LTE cell-specific reference signals~\cite{Book_LTE}) to combat the sporadic noises and ensure reliable measurement of CSI.
On the other hand, 
the time overhead of CSI acquisition should be significantly smaller than the channel coherence time, 
so that there is still a large proportion of time for data transmission before the measured CSI stales.

In vibration based MIMO over skin, 
unlike RF transceivers that can transit between Tx and Rx in $\mu s$, the motor consumes a much longer time to overcome the static inertia of the internal movable mass. Specifically, as measured in~\cite{roy2015ripple}, a motor needs 30ms to reach a stable vibration status from a cold start, refer to \textit{ramping effect}, and another 10ms to completely mute even a braking-voltage is applied, refer to \textit{ringing effect}. Hence, for a time division half-duplex system\footnote{Note that an alternative to time division half-duplex is frequency division half-duplex. However, skin MIMO in frequency division half-duplex is not practical because the guard band between the downlink and the uplink would be too small to mitigate the inter-link interference, e.g., at most in the order of several kHz. 
Also, the entire guard band would be wasted, leading to an unacceptably high overhead.}, the vibration sensors have to wait for 10ms when switching from Tx to Rx, to avoid the received signal been overwhelmed by the self transmitting signal. Furthermore, when switching from Rx to Tx, an even larger time period of 30ms would be wasted for a transmitting motor to enter a valid vibration status.

\begin{figure}[t]
	\centering
	\includegraphics[scale=0.68]{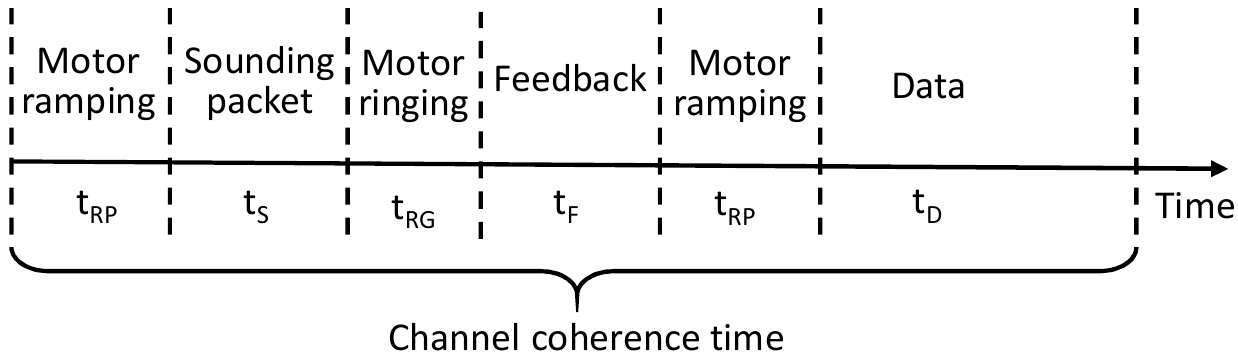}
	\caption{The challenge of using channel sounding based CSI acquisition in vibration MIMO. It is expected that $2t_{RP}+t_S+t_F+t_{RG}<<t_D$. But due to the ramping and ringing effect of motor, $2t_{RP}+t_{RG}$ already exceeds the channel coherence time in active scenarios, making sounding based CSI acquisition impractical. }
	\label{fig:challenge}
\end{figure}

As shown in Figure~\ref{fig:challenge}, 
if the motor wants to transmit some data, 
it first sends a sounding packet to initiate the CSI acquisition. 
Without considering the time overhead of the round trip (length of sounding packet plus uplink feedback time $t_S+t_F$), 
$2t_{RP}+t_{RG} = 70ms$ would be wasted before actual data transmission. 
For the scenarios with hand movement, 
such time already exceeds the channel coherence times of browsing (44ms) and typing (40ms), 
making such conventional CSI acquisition meaningless. 
Even for resting with a 150ms coherence time, more than 46\% time resource is already wasted. 
Moreover, 
if the motor is working on its resonant frequency (e.g., 200Hz for most off-the-shelf motors), 
a sounding packet merely containing 10 sine cycles leads to a time length of 50ms, 
which further slashes the remaining time for data transmission from 80ms to 30ms. 
As a result, 
conventional channel sounding based CSI acquisition is infeasible or inefficient in vibration MIMO communications. 


In specific scenarios where the channel changes rapidly, 
e.g., a fast-moving vehicle, 
conventional RF MIMO systems have to give up CSI acquisition and perform open-loop MIMO without the assistance of CSI. 
Specifically, 
an open-loop MIMO scheme uses a deterministic precoding matrix or a randomized sweeping of precoding matrices~\cite{Book_LTE}. 
A simple example of open-loop MIMO is an identity matrix with equivalent powers on all TX antennas. 
However, 
as we will present in Section~\ref{s:capacity}, open-loop MIMO suffers from a large capacity loss and achieves a data rate just slightly higher than that of SISO.

\begin{figure*}[t]
	\centering
	\includegraphics[scale=0.65]{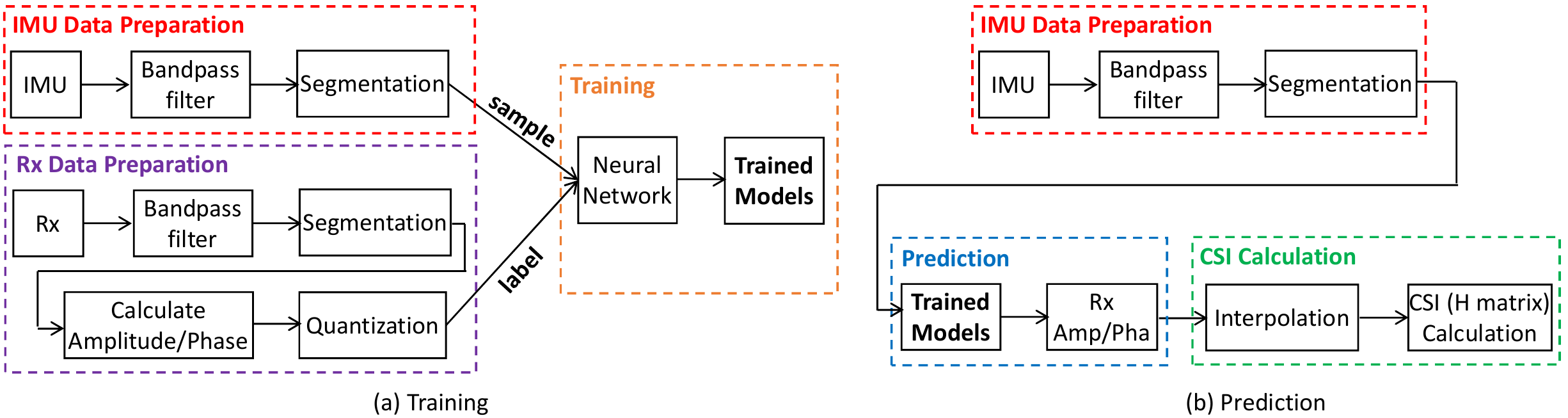}
	\caption{Off-line training and real-time prediction pipelines for Skin-MIMO. The bandpass filters extract IMU or receiver signals for specific pilot carriers.}
	\label{fig:framework}
\end{figure*}

\begin{figure}[t]
	\centering
	\includegraphics[scale=0.6]{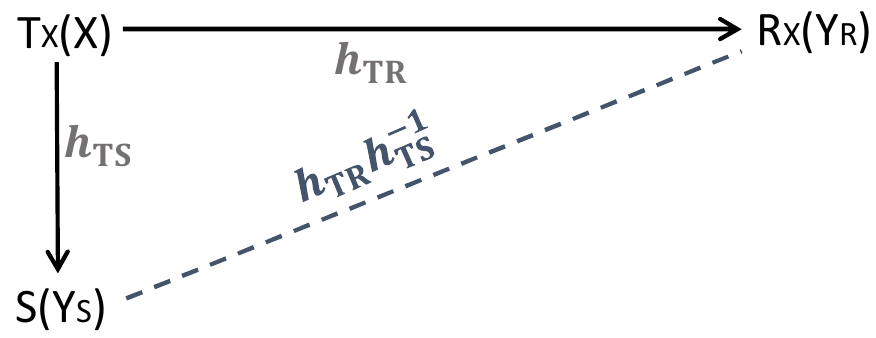}
	\caption{Communication theoretic illustration of how the vibration observed at a distant receiver is correlated with the vibrations measured by a sensor co-located with the transmitter.}
	\label{f:corr}
	\vspace{-0.2in}
\end{figure}

\section{Proposed deep learning-based CSI Acquisition }
\label{s:deep}

Our objective is to enable the transmitter to estimate CSI without having to send channel sounding packets and wait for CSI feedback from the receiver. We propose to achieve this by using vibration sensors, such as accelerometers and gyroscopes, fitted in the transmitter. We first show analytically that skin vibrations observed near the vibrating motor (transmitter) are correlated to the vibrations observed at a distant receiver. Then we propose to learn the Tx-Rx vibration correlations using deep neural networks, so the transmitter can predict the CSI observed at the receiver simply from the sensor samples available right inside the transmitter. Moreover, unlike conventional channel sounding based method that has to wait for a round trip of sounding packet and feedback, the proposed learning based approach can measure the CSI continuously. Specifically, the IMU (Inertial Measurement Unit) sensor data can be processed for prediction with an \textit{overlapped} sliding window so that the freshest (i.e., capturing the real channel more accurately) CSI can be obtained.


\subsection{Tx-Rx Vibration Correlation}

From communications theory points of view, Figure~\ref{f:corr} illustrates how a vibration signal (X) generated at a transmitter (Tx) is observed by a sensor (S) close to the transmitter as well as by a distant receiver (Rx). Let us denote the signals observed by the sensor and the receiver as $Y_{S}$ and  $Y_{R}$, respectively, and the channel response between Tx-Rx and Tx-S as $h_{TR}$ and $h_{TS}$, respectively. From the outcomes of frequency response investigations reported  in~\cite{zhang2017bioacoustics}, we can assume that skin vibration channels are linear. Then $Y_{S}$ and  $Y_{R}$ can be obtained as:
\begin{eqnarray}
Y_{S} & = & h_{TS}X+n_{S} \\ \nonumber
Y_{R} & = & h_{TR}X+n_{R}
\label{eq:1}
\end{eqnarray}
where $n_{S}$ and $n_{R}$ represent noise at the sensor and receiver, respectively. Ignoring the noise and applying some simple mathematical transformations, the vibration signal at the receiver can be expressed as a function of the signal observed by the sensor as follows:
\begin{equation}
Y_{R}= h_{TR}h_{TS}^{-1} Y_{S}.
\label{eq:corr}
\end{equation}

Equation~\ref{eq:corr} reveals that there exists a correlation between the vibration signal observed at the receiver and the vibrations measured by the sensor co-located at the transmitter, which is determined by the channel response functions between the transmitter and the receiver and between the transmitter and the sensor. While the correlation is understandably complex and difficult to derive mathematically, it could be potentially \textit{learned} if we had enough training data available. If we can learn the correlation, we can use the trained model to predict or infer the observed CSI at the receiver simply from the sensor samples available right at the transmitter. In the rest of this section, we present our proposed CSI learning and estimation approach using deep neural networks.


\subsection{CSI Learning and Prediction Framework}

Following the popular OFDM physical layer concept, we assume that the total vibration spectrum is divided into a large set of data carriers and a small set of pilot carriers. The usual goal of the pilot carriers would be to help receivers demodulate the symbols accurately, but in Skin-MIMO, we exploit them further for CSI prediction at the transmitter. 

We propose that different transmitting antennas transmit a different set of pilots, which allows the receivers to identify the transmit antenna of a receiving pilot. The set of all pilot carriers are distributed across the whole vibration spectrum to be used, so we can limit the learning and prediction only to the limited number of pilot carriers and use interpolation to predict CSI for any arbitrary data carriers within the spectrum.    

Figure~\ref{fig:framework} illustrates the off-line training and real-time prediction pipelines for Skin-MIMO. We utilize a bandpass filter to separate each pilot carrier and calculate the amplitude and phase of a segment so that current CSI can be acquired. The goal in the training phase is to train a neural network so, for each pilot carrier, it can predict or classify the quantized amplitude and phase values at each of the receiving antennas from the IMU samples observed at the transmitter. The amplitude and phase values for data carriers are then derived through interpolation and finally the CSI matrix is obtained from them. Although the proposed method is able to acquire CSI without channel probing, it incurs a burden on the user to pre-train the model before real MIMO communication, which we will discuss in Section~\ref{s:future_work}.

In real-time, the trained models require some time to predict corresponding quantization levels, refers to as \textit{inference time}. However, as long as the inference time is within the channel coherence time, the predicted CSI is valid. This is because the proposed framework enables continuous CSI acquisition by overlapping the sampled IMU data. Ideally, once an IMU sample is collected, a fresh CSI can be predicted by reusing the IMU data within a sliding window.

\subsection{Neural Network Design}
\begin{figure}[t]
	\centering
	\includegraphics[scale=0.61]{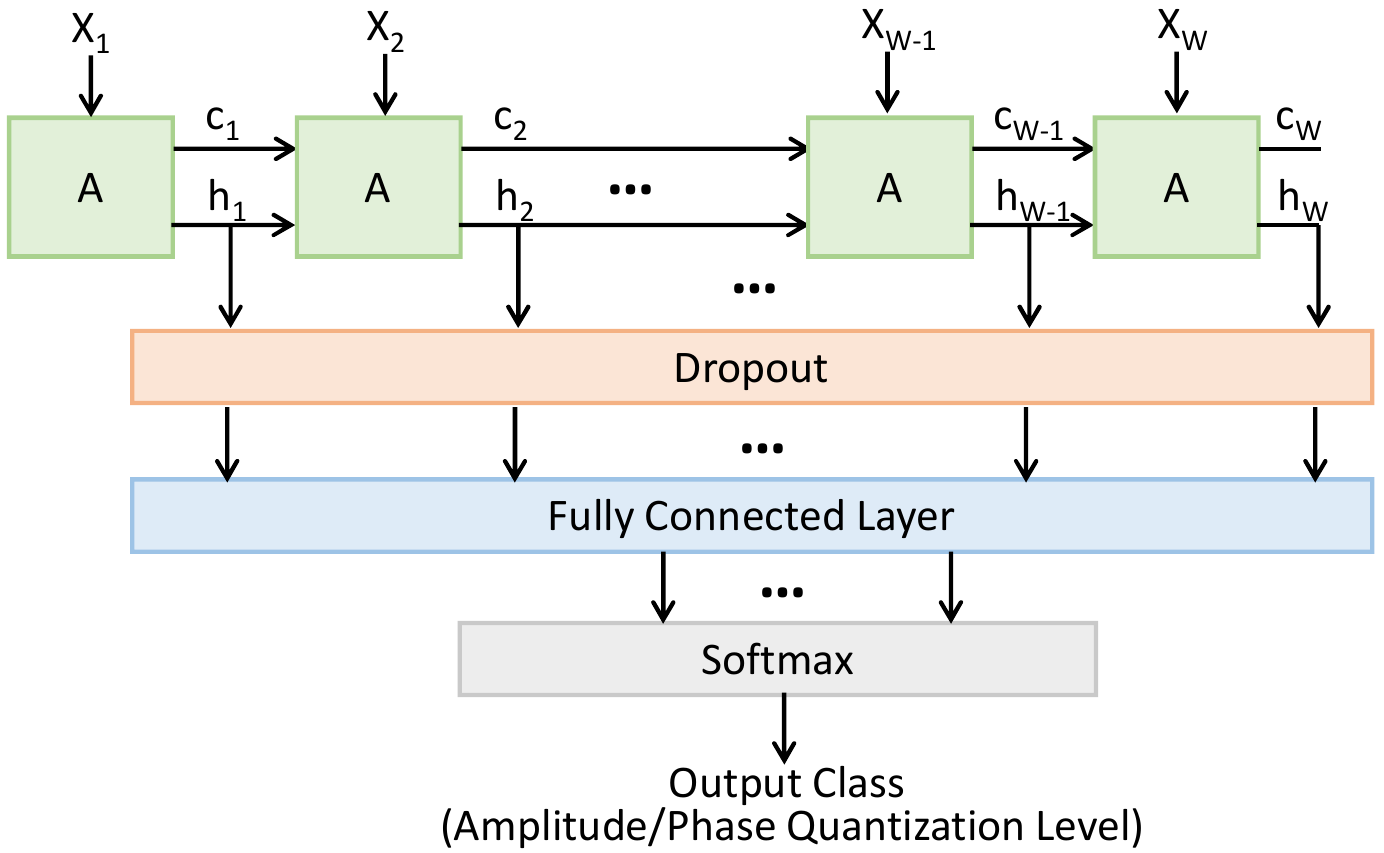}
	\caption{The structure of the proposed LSTM network.}
	\label{fig:deep_neural_network}	
\end{figure}
The goal of the neural network is to learn the correlation between the transmitter sensor (IMU) data and the pilot signals received at the receiving antennas, which is defined by a complicated channel function involving two channels as shown in Figure \ref{f:corr} . Due to the time variation of the channels, there exists abundant temporal information in the collected signals. Thus, we propose to use Long short-term memory (LSTM) neural networks as the classifier to predict the received quantized values of signal amplitude and phase. LSTM has been extensively exploited in various applications and recognized as a superior choice for time series data classification and forecasting~\cite{truong2018capband,zhang2018internet,plotz2018deep,hochreiter1997long}. 

For each pilot carrier, we need to classify both amplitude and phase quantization levels at each receiving antenna. For $Z$ pilot carriers and $N$ receive antennas, we therefore propose to train a total of $Z \times N$ LSTMs to detect amplitudes and another $Z \times N$ LSTMs to classify phase quantizations.

Figure~\ref{fig:deep_neural_network} shows the proposed structure of each LSTM. In the top layer, each green block (denoted as A) represents an LSTM cell, which takes an L-dimension sensor vector data as the input. For example, if we wish to use only a 3-axis accelerometer ($L=3$) for CSI prediction, then each LSTM cell takes a 3-axis acceleration sample as input. For a W-sample segmented time-series of sensor samples, we have W LSTM cells at the input layer.

An LSTM cell learns information from its input and produces an output $h_i$. Meanwhile, it remembers some message (termed as cell state $c_i$) that might be useful later and passes $c_i$ to the subsequent cell. It is the cell state mechanism that enables LSTM neural networks to explore temporal correlations in sequential signals. Each cell consists of 128 hidden neurons. Then, we add a dropout layer with a probability of 0.5 to avoid overfitting~\cite{srivastava2014dropout}. A fully connected layer is used to generate the probability of current sample on each class and then a Softmax layer is responsible to produce the final result by selecting the class with the highest probability.






\begin{figure}[t]
	\centering
	\includegraphics[scale=0.61]{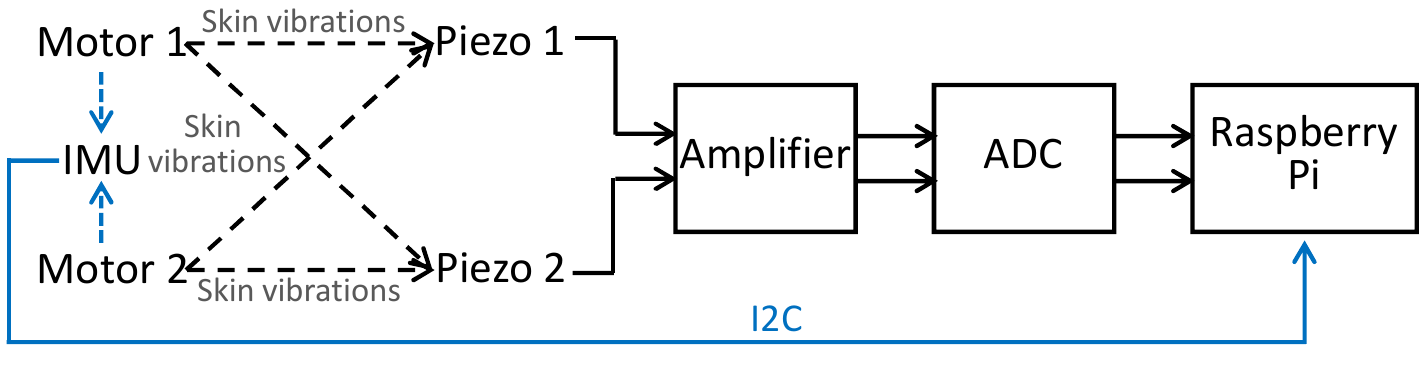}
	\caption{System diagram of the $2 \times 2$ MIMO testbed fitted with IMU. Because the IMU is a digital MPU9250, instead of ADC, it utilizes I2C bus for connecting to the Raspberry Pi.}
	\label{f:imu_setup}
\end{figure}

\section{Performance Evaluation of Skin-MIMO}
\label{s:evaluation}
\subsection{Testbed and Data Collection}
To evaluate the performance of Skin-MIMO, we extend the $2 \times 2$ MIMO testbed in Figure \ref{fig:testbed} by integrating an IMU between the two motors as shown in Figure \ref{f:imu_setup}. For the IMU, we sample both the 3-axis accelerometer and the 3-axis gyroscope, which will allow us to compare the performance of linear and angular motion sensors in predicting CSI values. We consider a total of 6 different pilot carriers with Motor1 transmitting on carriers 90Hz, 110Hz, and 130Hz, and Motor2 transmitting on carriers 100Hz, 120Hz, and 140Hz. 

We collect data from the left hands of two real subjects for a total of five minutes for each subject with some breaks after each minute to avoid heating up the motors from continuous operations\footnote{Note that in real life, motors are expected to operate only intermittently and turn off when there are no data transmissions.}. The six-axis IMU at the wrist and the two piezo transducers (receivers) at the fingers are sampled simultaneously at 300Hz using the same Raspberry Pi clock. Thus, for each subject and for each the 6 pilot carriers: (i) we have a total of 90,000 6-axis IMU samples for each of the two receivers, and (ii) by using a 100ms signal window (30 samples) to calculate the amplitude and phase of the received signals, we obtain a total of 3,000 amplitude/phase per receiver. 

\subsection{LSTM implementation, training, and testing}

For each of the 6 pilot carriers, we implement 2 LSTMs, one for each receiver, for detecting amplitudes and similarly 2 LSTMs for phases. For each of these LSTMs, we created 3 \textit{IMU versions}, one for gyro only(GYRO), one for accelerometer only (ACC), and one for all 6-axis IMU data (ACC+GYRO). All LSTMs are implemented in Keras based on Python and executed with GeForce RTX 2080 Ti Graphics Card from NVIDIA~\cite{gpu}. Using experimental data, we tune the different parameters used to train the LSTM models and obtain the following optimal settings: batch size is 32, objective loss function is cross-entropy, and optimizer is Adam~\cite{kingma2014adam}. Both training and testing are subject-dependent, but for a given subject, all testing is performed using 10-fold cross-validation.


\subsection{Performance Metrics}
We study the following performance metrics:
\subsubsection{Prediction Accuracy as TPR}
Given an IMU sample, each trained model will predict the quantization level of amplitude or phase. True positive rate (TPR) refers to the percentage of the tests that correctly predict the quantized level. 
\subsubsection{Prediction Accuracy as RMSE}
 TPR only captures the quality of the LSTMs in correctly predicting the quantization levels of amplitudes/phases, but does not provide any clue about the difference between the true value and the predicted value. We use the root mean square error (RMSE) to study such performance, which is computed as: 
\[
    RMSE = \sqrt{\frac{\sum_{1}^{N}(X(G_{k})-X(P_{k}))^2}{N}}
\]

where $G_{k}$ and $P_{k}$ are the ground truth and predicted quantization levels for the $k_{th}$ test sample (out of $N$ samples), respectively, and $X$ is a dictionary recording corresponding exact ground truth amplitude/phase values for each quantization level.

\subsubsection{MIMO Capacity}
MIMO capacity is defined as the data rate with a bandwidth of 1Hz, i.e., bit/s/Hz. Given a ground truth CSI (perfect) $\mathbf{H}$ and a measured CSI (imperfect) $\hat{\mathbf{H}}$, one is able to calculate the effective capacity using singular value decomposition (SVD)~\cite{bizaki2011mimo}. Firstly, we apply the SVD to $\hat{\mathbf{H}}$, so
\begin{equation}
    \hat{\mathbf{H}}=\hat{\mathbf{U}}\hat{\mathbf{\Lambda}}\hat{\mathbf{V}}^{H},
    \label{eq:4}
\end{equation}
 where $H$ is the hermitian transpose of a matrix, and $\hat{\mathbf{U}}$ and $\hat{\mathbf{V}}$ are two unitary matrix. Then, let $\mathbf{\Gamma} = \hat{\mathbf{U}}^{H}$ represents the decoding matrix at the receiver, and $\mathbf{W} = \hat{\mathbf{V}}$ represents the precoding matrix at the transmitter. The received signal $\mathbf{y}$ can be written as
\begin{equation}
    \mathbf{y=\Gamma H W x + n = \hat{U}}^{H} \mathbf{H \hat{V} x + n},
    \label{eq:5}
\end{equation}
    
where $\mathbf{x}$ is the transmitted signal and $\mathbf{n}$ is the noise. Let 

\begin{equation}
    \mathbf{\Phi = \hat{U}^{H} H \hat{V}} = 
    \begin{bmatrix} 
 {\phi}_{11} & {\phi}_{12} \\
 \phi_{21} & \phi_{22}, 
 \end{bmatrix}
\end{equation}
then, for $2\times 2$ MIMO, the signal-to-interference-plus-noise ratio can be calculated as $SINR_1 = \frac{\phi_{11}^2SNR}{\phi_{12}^2SNR+1}$, $SINR_2 = \frac{\phi_{22}^2SNR}{\phi_{21}^2SNR+1}$. Note that we collect the piezo signals when both motors are muted as the noise, and calculate SNR using signal power divided by noise power. The effective capacity can be calculated as 
\begin{equation}
    C = log_2(1+SINR_1)+log_2(1+SINR_2)
    \label{eq:cap}
\end{equation}

In open-loop MIMO (OL-MIMO), the knowledge of CSI (i.e., $\hat{\mathbf{H}}$) is not available, so the precoding matrix $W$ is set to a $2\times 2$ identity matrix.


\begin{figure}[t]
	\centering
	\includegraphics[scale=0.62]{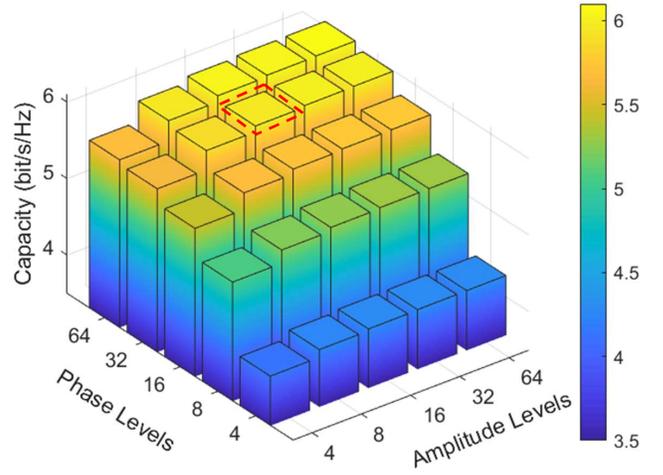}
	\caption{Impact of quantization levels on capacity, which indicates that increasing the quantization level of phase can result in more capacity improvement compared to amplitude.}
	\label{fig:quanti_level_impact}	
	\vspace{-0.2in}
\end{figure}

\subsection{Impact of Quantization Levels}

To study the effect of quantization error, in Figure~\ref{fig:quanti_level_impact}, we plot the MIMO capacity as a function of quantization levels for the Oracle Skin-MIMO, i.e., when the LSTMs predict quantization levels with 100\% accuracy (TPR = 100\%). The interesting observation is that we are able to increase capacity more significantly by increasing the number of quantization levels in phase compared to amplitude. This could be explained by the fact that the range of phase shifts is rather large (180 degrees) compared to amplitude, which is only a few millivolts. From  Figure~\ref{fig:quanti_level_impact}, we find that 16 amplitude levels and 32 phase levels increase MIMO capacity significantly, but increasing the number of quantization levels any further does not provide any further major improvement. Since increasing the number of quantization levels increases LSTM complexity (more classes to detect), the rest of the performance analysis is carried out using 16 amplitude and 32 phase levels. 



\subsection{LSTM Prediction Performance}


Table~\ref{tab:prediction_performance} compares TPR and RMSE performances for three versions of the sensor --- GYRO only, ACC only, and ACC fused with GYRO (6 IMU axes). Phases are in radian and each value in the table is averaged by both received signals at six frequencies. We can see that 90\% TPR can be achieved. In terms of RMSE, the range of amplitude for subject 1 and subject 2 are 0.114 and 0.035 respectively, and those for phase are both $2\pi$. In the best case (GYRO+ACC), the RMSE for amplitude and phase is within 8\% and 7\%, respectively, for both subjects, which suggests that we obtain a very accurate prediction of the CSI.

\begin{figure}[t]
	\centering
	\includegraphics[scale=0.45]{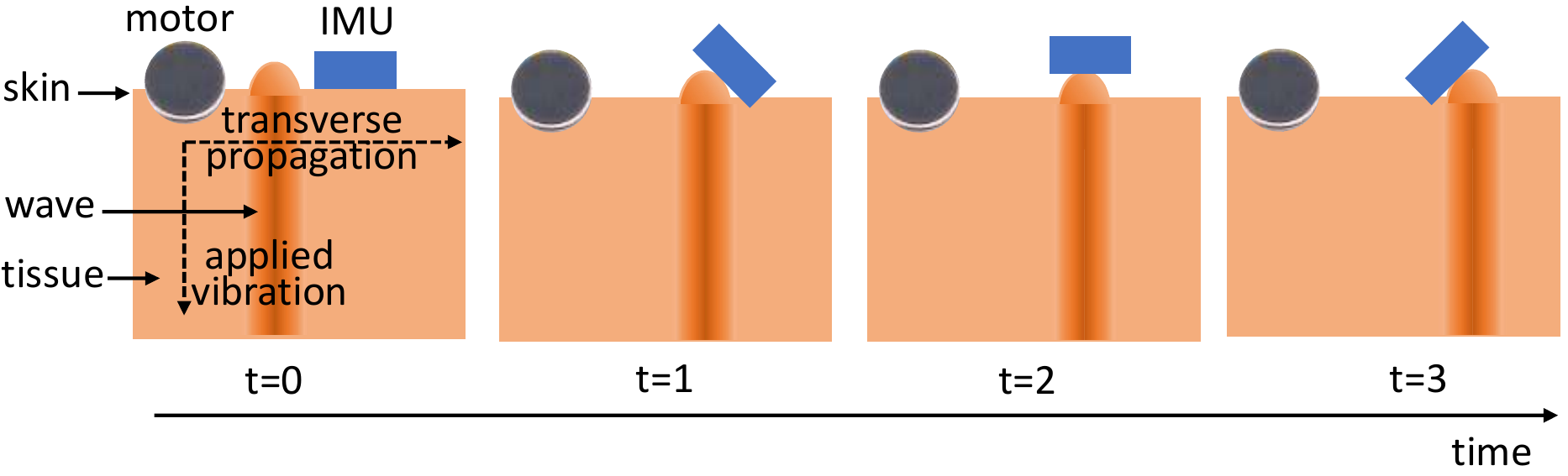}
	\caption{Illustration of transverse propagation of vibration (vertically applied vibration causing horizontal wave motion) and how the soft human skin is deformed to create rotational movement of IMU. }
	\label{fig:transverse_propogation}	
\end{figure}

 An interesting finding is that gyroscope achieves significantly higher TPR and lower RMSE compared to accelerometer, while sensor fusion between these two motion sensors does not provide a noticeable performance improvement. This can be perhaps explained based on the ways vibrations propagate over human skin. As illustrated in Figure~\ref{fig:transverse_propogation}, although the motors in our testbed apply vibrations perpendicularly to the skin, these vibrations propagate on the skin \textit{transversely}~\cite{harrison2010skinput} and unlike solid objects, human skin is soft and can be deformed by these transverse vibration waves. As a result, the crest and trough of sine waves can tilt the sensor creating an angular response, which can be better captured by a gyroscope, because it measures \textit{angular} velocity unlike the accelerometer which measures \textit{linear} acceleration. 

\begin{table}[t]
\centering
\caption{LSTM prediction performance: GYRO vs. ACC vs. GYRO+ACC.}
\label{tab:prediction_performance}
\ra{1.2}
 \setlength\tabcolsep{6.6pt}
\begin{tabular}{crrrrr}\toprule
\multicolumn{2}{c}{\textbf{}} && \multirow{2}{*}{\textbf{GYRO}} & \multirow{2}{*}{\textbf{ACC}} & \textbf{ACC+} \\ 
 &&  &  && \textbf{GYRO} \\ \midrule
 
\multirow{7}{*}{\textbf{TPR}}& \multirow{2}{*}{\textbf{Subject1}}& amplitude & 88.61\% &  69.54\% & 88.55\% \\
 & & phase & 89.88\%  &  74.97 \%& 89.66\% \\ \cmidrule{2-6}
 & \multirow{2}{*}{\textbf{Subject2}}&  amplitude & 90.29\% & 77.68 \% & 90.37\% \\
 & & phase & 90.74\%  &   81.86\%& 90.44\% \\ \cmidrule{2-6}
 
 & \multirow{2}{*}{\textbf{Average}}&  amplitude & 89.45\% & 73.61 \% & 89.46\% \\
 & & phase & 90.31\%  &   78.42\%& 90.05\% \\ \hline

 \multirow{7}{*}{\textbf{RMSE}}& \multirow{2}{*}{\textbf{Subject1}} & amplitude & 0.011 & 0.018  & 0.010 \\
  & & phase & 0.458  & 0.737  & 0.447 \\ \cmidrule{2-6}
  & \multirow{2}{*}{\textbf{Subject2}}&  amplitude & 0.003 &0.005   & 0.003 \\
 & & phase &  0.409 &  0.607  & 0.407 \\ \cmidrule{2-6}
 
 & \multirow{2}{*}{\textbf{Average}}&  amplitude & 0.007 &0.011   & 0.007 \\
 & & phase &  0.434 &  0.672  & 0.427 \\ 
\bottomrule
\end{tabular}
\end{table}


\subsection{Analysis of MIMO Capacity}
\label{s:capacity}
\begin{figure}[t]
	\centering
	\includegraphics[scale=0.68]{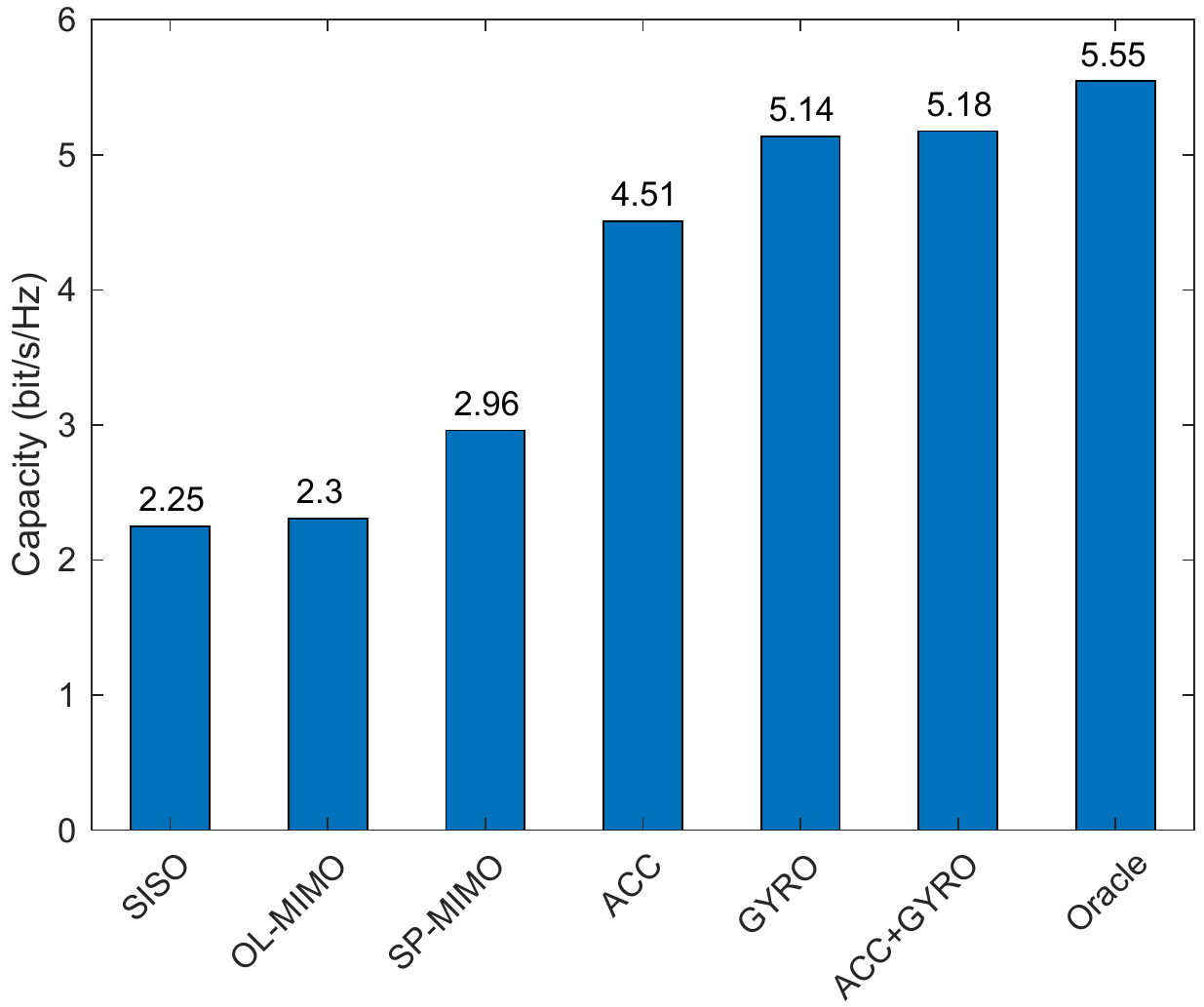}
	\caption{Comparison of capacity under different schemes (averaged by two subjects).}
	\label{fig:capacity_com}
	\vspace{-0.2in}
\end{figure}

After predicting the amplitude and phase of 6 pilot carriers, we derive the amplitude and phase of data carriers at different frequencies through spline interpolation, which is a widely used interpolation technique~\cite{spath1995one}. Using the interpolated amplitudes and phases, we are able to calculate the channel response of each of the four channels for these data carriers using Equation~\ref{eq:csi}, and thus estimate the channel matrix $\mathbf{\hat{H}}$. Then, using Equations~(\ref{eq:4})-(\ref{eq:cap}), we compute the MIMO capacity. SISO capacity is obtained as the mean of the four individual subchannels (4 subchannels in $2\times 2$ MIMO) capacities.  {\bf Note that, when calculating MIMO capacity, the SINR is divided by 2 (i.e., number of transmission antennas) to ensure the transmission power is equivalent to that of SISO.} For sounding packet based MIMO (SP-MIMO), we consider a time overhead of 80ms (30ms ramping effect plus 50ms SP length) over 150ms channel coherence time. Then, the capacity is calculated via the oracle capacity multiplied by the proportion of remaining time for data transmission.


Figure~\ref{fig:capacity_com} compares the capacity, averaged over the two subjects, under different MIMO schemes for data carrier 125Hz\footnote{Although the absolute capacities are slightly different for different data carriers, the comparison among the six MIMO schemes exhibits similar trends.}. It is clear that without the knowledge of CSI, OL-MIMO only achieves a comparable capacity with SISO, thereby losing the spatial multiplexing gain of MIMO. The Oracle capacity refers to the capacity of the proposed LSTM-based Skin-MIMO assuming perfect classification of the amplitude and phase quantization levels. We can see that GYRO-based CSI prediction yields higher MIMO capacity compared to ACC-based prediction, which is consistent with the LSTM prediction performance results analyzed earlier in Table~\ref{tab:prediction_performance}. Overall, the proposed deep learning based CSI acquisition can achieve 93\% of the Oracle capacity, and increase the vibration communication capacity by a factor of 1.8 and 2.3 compared to SP-MIMO and SISO/OL-MIMO, respectively.
\vspace{0.1in}

\section{Related work}
Vibration communication exploits the propagation of vibrations over a medium for information transmission. Generally, a vibration communication system consists of a vibrator (e.g., vibratory motors and bone conductors), a vibration sensor (e.g., accelerometers, microphones, and piezo transducers), and a medium (e.g., air, water, and solid objects). 


Prior works have successfully demonstrated the feasibility of using vibratory motor and accelerometer to communicate through physical objects within a short distance~\cite{yonezawa2015vinteraction,kim2015vibration,yonezawa2011vib,yonezawa2011vib,roy2015ripple,roy2015ripple}. Yonezawa et al. presented VibConnect, a mobile system which aims to transfer URL information from smartphones to laptops through vibration signal~\cite{yonezawa2011vib}. Using OOK (On-Off Keying) modulation, VibConnect achieves up to 10 bit/s data rate when the motor and accelerometer are directly touched. With the same transmitter and receiver, the following works improve the performance either by extending the communication distance or enhancing the data rate. Hwang et al. proposed VibeComm that extends the distance to 50cm while still achieving similar data rate~\cite{hwang2012privacy}. With in-touch communication, Romit et al. presented Ripple, which significantly lifts the data rate up to 200 bit/s by using multi-carrier modulation~\cite{roy2015ripple}. Furthermore, Ripple II leveraged microphone (can enable much wider bandwidth due to high sampling rate) as the receiver and applied OFDM modulation to dramatically boost the data rate to 30 kbit/s~\cite{roy2016ripple}. 

Using skin as the medium, Zhang et al. investigated vibration communication over human body~\cite{zhang2017bioacoustics}. Specifically, using a bone conductor as the transmitter and an/a accelerometer/microphone as the receiver, they analyzed the characteristics of human body channel, such as frequency response and path loss, and demonstrated several practical applications such as exchanging personal information during handshaking. Modulating vibrations with FSK (frequency shift keying), they achieved a data rate of up to 105 bit/s with 38cm distance over human arm. Similarly, Ripple II transmits vibrations using a finger ring fitted with a motor to a touched object and achieves a data rate of 7.41 kbit/s with OFDM.

Current research on vibration communication systems over the human body are all based on SISO designs. Exploiting MIMO to boost data rates remains an open problem. We observe that the complex bone and muscle structure of the human
body has the potential to create uncorrelated channels and therefore enable vibration MIMO over human skin. In addition, we identified that gyroscope is a superior sensor in detecting skin vibrations, which allows more design choices for future vibration communication on soft materials.

\vspace{0.2in}
\section{Conclusion}

In this paper, we investigated MIMO vibration communication over human skin. Firstly, we have demonstrated the feasibility of vibration MIMO between wrist and fingers, where the existence of complex bone and muscle structure helps achieve path diversity for vibration signals. Secondly, based on the observation that vibration signal from a position co-located with the transmitter is correlated with that from the receiver, we have proposed a deep learning based CSI acquisition framework to allow the transmitter to obtain real-time CSI simply from accelerometer or gyroscope samples without using conventional sounding packets. Lastly, with a $2\times 2$ vibration MIMO testbed built in our laboratory, we have collected skin vibration data from two subjects and evaluated the performance of the proposed CSI acquisition framework. The experimental results have demonstrated that our method can improve the capacity of a $2\times 2$ MIMO  by $2.3\times$ compared to SISO and open-loop MIMO.


\section{Future Work}
\label{s:future_work}
The current work should be considered as a first attempt to understand the potential of vibration-based MIMO communications over human skin. Much work remains before the concept of Skin-MIMO can be ready for deployment in commercial wearable products. Here, we discuss some of the important future research directions:
\begin{itemize}
    \item {\bf Reduce training time with transfer learning:} The need for pre-training the CSI learning model is an overhead of the proposed Skin-MIMO system. It is therefore important to find solutions that can reduce this training time to a level that is acceptable by the users. Transfer learning \cite{weiss2016survey,zoph2016transfer
} is a concept in machine learning, which facilitates transfer of knowledge from one model to another. Thus, transfer learning can be used to reduce training time of Skin-MIMO as follows. For example, a manufacturer of a wearable device can collect data from a large number of subjects and train a generic learning model, which is released with the product. The generic model, which could be trained for a specific age group, contains some knowledge that is applicable for any person in that group, but it would perform poorly on any specific person due to lack of some detailed information only relevant for that person. Thus, with minimum extra training, the generic model can be personalized for anyone saving significant training time compared to the case when no transfer learning is applied and the model has to be learned from scratch. This future work would focus on designing novel transfer learning models and algorithms that can successfully reduce training time for predicting skin vibration channels. 
 
    \item {\bf Minimize deep learning inference time with model pruning:} Deep learning models contain large number of parameters with floating point weights, which increases their inference time, i.e., the time it takes for a trained model to predict the target class. The capacity results discussed in this paper for Skin-MIMO assumed that inference time is smaller than the channel coherence time. We ported the trained LSTM models to an Apple iPhone X running iOS 12.1.3 and test with real IMU samples. We found that it takes approximately 38ms to predict the amplitude or phase quantization level for all the pilot carriers. This means that on a more resource-constrained wearable device, it could take longer to predict CSI and violate the channel coherence time limit. A future research direction, therefore, would be to focus on compressing the deep learning model using various techniques such as structured pruning~\cite{han2015deep}, deep compression~\cite{han2015deep}, and evolutionary  pruning~\cite{yang2012evolutionary}. Although this is a very challenging task because the pruning must be achieved without reducing the prediction accuracy of the model too much, the encouragement comes from a most recent study of Google demonstrating that the inference time of face recognition can be as low as 0.6ms in commercial mobile phones~\cite{bazarevsky2019blazeface}.  
\end{itemize}


\balance

\bibliographystyle{IEEEtran} 
\bibliography{infocom_ref}

\end{document}